\documentclass[useAMS]{mn2e}
\usepackage{natbib}
\usepackage{graphicx}
\usepackage{gensymb}
\usepackage{footnote}
\begin{document}
\title{\textit{Kepler K2} Observations of the Intermediate Polar
FO Aquarii}
\author[M. R. Kennedy et al.]
{M. R. Kennedy$^{1,2}$\thanks{Email: markkennedy@umail.ucc.ie}, P. Garnavich$^2$, E. Breedt$^3$, T. R. Marsh$^3$, B. T. G\"{a}nsicke$^3$,\newauthor D. Steeghs$^3$, P. Szkody$^4$, Z. Dai$^{5,6}$ \\$^1$Department of Physics, University College Cork, Ireland.\\$^2$Department of Physics, University of Notre Dame, Notre Dame, IN 46556.\\$^3$Department of Physics, University of Warwick, Gibbet Hill Road, Coventry, CV4 7AL, UK\\$^4$Department of Astronomy, University of Washington, Seattle, WA, USA\\$^5$Yunnan Observatories, Chinese Academy of Science, 650011, P. R. China \\$^6$Key Laboratory for the Structure and Evolution of Celestial Objects, Chinese Academy of Sciences, P. R. China}
\maketitle
\date{}
\pagerange{\pageref{firstpage}--\pageref{lastpage}} \pubyear{}
\label{firstpage}

\begin{abstract}
We present photometry of the intermediate polar FO Aquarii obtained as part of the \textit{K2} mission using the \textit{Kepler} space telescope. The amplitude spectrum of the data confirms the orbital period of 4.8508(4) h, and the shape of the light curve is consistent with the outer edge of the accretion disk being eclipsed when folded on this period. The average flux of FO Aquarii changed during the observations, suggesting a change in the mass accretion rate. There is no evidence in the amplitude spectrum of a longer period that would suggest disk precession. The amplitude spectrum also shows the white dwarf spin period of 1254.3401(4) s, the beat period of 1351.329(2) s, and 31 other spin and orbital harmonics. The detected period is longer than the last reported period of 1254.284(16) s, suggesting that FO Aqr is now spinning down, and has a positive $\dot{P}$. There is no detectable variation in the spin period over the course of the \textit{K2} observations, but the phase of the spin cycle is correlated with the system brightness. We also find the amplitude of the beat signal is correlated with the system brightness.
\end{abstract}


\section{Introduction}
FO Aquarii (FO Aqr) is a binary system containing a magnetic white dwarf (WD) which accretes material from a near-main sequence companion. It was initially discovered as the X-ray source H2215-086 \citep{Marshall1979} and classified as a cataclysmic variable upon discovery of an optical counterpart by \cite{Patterson1983}. It is considered the King of the intermediate polar class of cataclysmic variables mainly due to the large amplitude (0.2 mag) of its spin signal \citep{Patterson1983}.

In intermediate polars (IPs), also called DQ Her systems after the first known member of this category, the white dwarf's magnetic field is high enough ($\sim$1-10 MG) that the material from the donor star is not accreted through a typical accretion disk extending close to
the white dwarf. There are three current models for accretion in an IP, with the applicable model for any given IP depending on the magnetic moment of the WD and the mass accretion rate. The first is the disk-fed model, where an accretion disk is present in the system, and the inner edge of the accretion disk is truncated by the magnetosphere of the WD, where the magnetic pressure of the WD exceeds the ram pressure of the disk. Material then follows ``accretion curtains'' to the magnetic poles of the WD \citep{Rosen1988}. The second accretion model is the stream-fed model, which has no accretion disk due to the magnetic field of the WD. In this model, the accretion stream entering the Roche lobe of the WD does not form a disk, but instead directly impacts the magnetosphere of the WD \citep{Hameury1986}. The third model is a combination of the first two models, and is known as the ``disk-overflow" model (\citealt{Lubow1989}; \citealt{Armitage1996}). In this model, material accretes via the ``accretion curtains'' of the disk fed model, but accretion can also occur from material in the accretion stream being caught by the magnetic field at the hot spot, following the stream-fed model.

In all of the above models, the EUV and X-ray light created by the impact of material onto the WD  is partially reprocessed into the optical by the accretion curtain, accretion disk or accretion stream. In the disk-fed and disk-overflow models, the material impacts onto both magnetic poles, while in the stream-fed model it impacts a single pole, and as these regions move into and out of our plane of sight, they give rise to modulations in the optical light curve. This allows for the detection of the white dwarf spin period $\left(2\pi/\omega\right)$ at optical wavelengths (for a review, see \citealt{Patterson1994}).

Understanding accretion in binary systems is not only important for CVs, but also for neutron star binaries, where the neutron star primary is spun up to have a period of milliseconds by accretion of matter \citep{Wijnands1998} and where the accretion disk can also be truncated by the strong magnetic moment of the neutron star. It is also important for understanding accretion in young stellar objects, low mass black hole binaries and active galactic nuclei, where the primary accretor is a supermassive black hole \citep{Scaringi2015}. Recent work has shown that, despite the differences in the primary star and the energy of these systems, there is a possible common link in their accretion physics. The rms-flux relation, which shows a linear relationship between the flux and amount of ``flickering'' in a system, seems to hold true for all of these systems (\citealt{Scaringi2015} and references therein). It is best seen as a break in the amplitude spectrum of an accreting system, and for CVs, this break occurs at $\sim 10^{-3}$ Hz (\citealt{Scaringi2012}; \citealt{Scaringi2015}).

The spin period of the white dwarf primary in FO Aqr was found to be 1254 s (\citealt{Shafter1982}; \citealt{Patterson1983}) while optical modulations of the light curve revealed periods of 4.03 h (\citealt{Shafter1982}; \citealt{Patterson1983}) and 4.85 h \citep{Patterson1983}. The 4.85 h signal was later confirmed as the orbital period $\left(2\pi/\Omega\right)$ of the system by \cite{Osborne1989}. Modulations at these wavelengths have also been seen in the X-ray and infrared. The beat period ($P_{B}=2\pi/\omega-2\pi/\Omega$) of 1350 s has also been seen in the optical and X-ray. 4 other harmonics of the spin frequency ($2\omega$ to $5\omega$) and various sideband periods have been seen in the X-ray \citep{Beardmore1998}.

The spin period in FO Aqr has fluctuated significantly since it was discovered. In 1982, \cite{Shafter1982} found it to be 1254.44(1) s, while during the 1980's the spin period was
refined to 1254.4514(3) s and had a positive $\dot{P}$ \citep{Shafter1987}, meaning the WD was spinning down. However, observations reported by \cite{Steiman-Cameron1989} from 1987 showed the period to be 1254.4511(4) s with a $\dot{P}\approx0$. In 1993, \cite{Kruszewski1993} found the spin period to be 1254.4518(2) s. This period had decreased to 1254.4446(2) s by 1998 (\citealt{Patterson1998}; \citealt{Kruszewski1998}) with a negative $\dot{P}$ reported, suggesting the WD was now spinning up. The period had further decreased to 1254.4441(1) s by 2003 \citep{Williams2003}, and to 1254.284(16) s by 2004 \citep{Andronov2005}.

There is strong evidence for disk-overflow accretion in FO Aqr. The strength of the spin frequency signal seen in the X-ray and optical amplitude spectrum supports the disk-fed model, while the strength of the beat frequency in the amplitude spectrum supports the stream-fed model. The combination of both of these suggest that FO Aqr accretes via the disk-overflow model (\citealt{Norton1992}; \citealt{Hellier1993}).

Here, we present photometry taken of FO Aqr as part of the \textit{K2} mission \citep{Howell2014}, spanning a range of 69 days with a cadence of 1 minute. We analyse the amplitude spectrum, and track the changing amplitudes of the various components that make up the spin modulation, and relate these changes to the changing accretion rate and modes in the system.

\section{Observations}
Observations of FO Aqr were taken by the \textit{Kepler} space telescope as part of the \textit{K2} mission observing campaign field 3 between 2014 Nov 14 and 2015 Feb 3 (MJD 56977-57046). The data are stored under the Ecliptic Plane Input Catalogue (EPIC) identification number EPIC206292760. The 1 min cadence data were extracted using PyKE \citep{2012ascl.soft08004S}. \textit{Kepler} fires its thrusters every 48 h, and as such, the target flux of FO Aqr decreased to zero while the spacecraft readjusted. Data taken during the thruster firing had \texttt{QUALITY} tags in the \textit{Kepler} FITS file data structure greater than 0, and were removed from the light curve.

\section{Light Curve Analysis}
\subsection{Full Light Curve}
\begin{figure}
	\includegraphics[width=80mm]{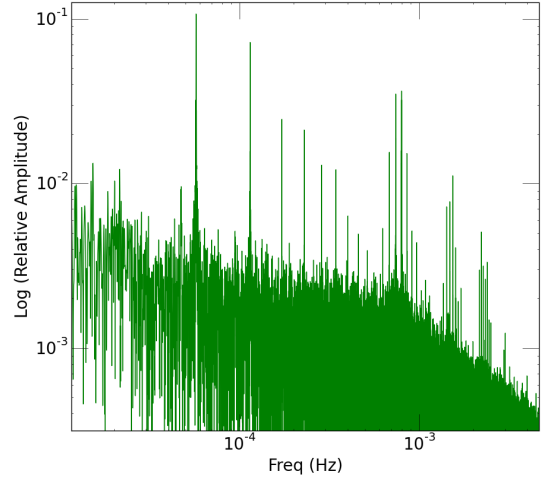}
	\caption{The amplitude spectrum of FQ Aqr. The break in the amplitude spectrum of FO Aqr occurs just before $10^{-3}$ Hz, similar to the break seen in other CVs such as MV Lyrae \citep{Scaringi2012}. The strongest signals are identified in Table \ref{frequency_tab} and shown in greater detail in Figure \ref{1minps}.}
	\label{pssingle}
\end{figure}

\begin{figure*}
	\includegraphics[width=180mm]{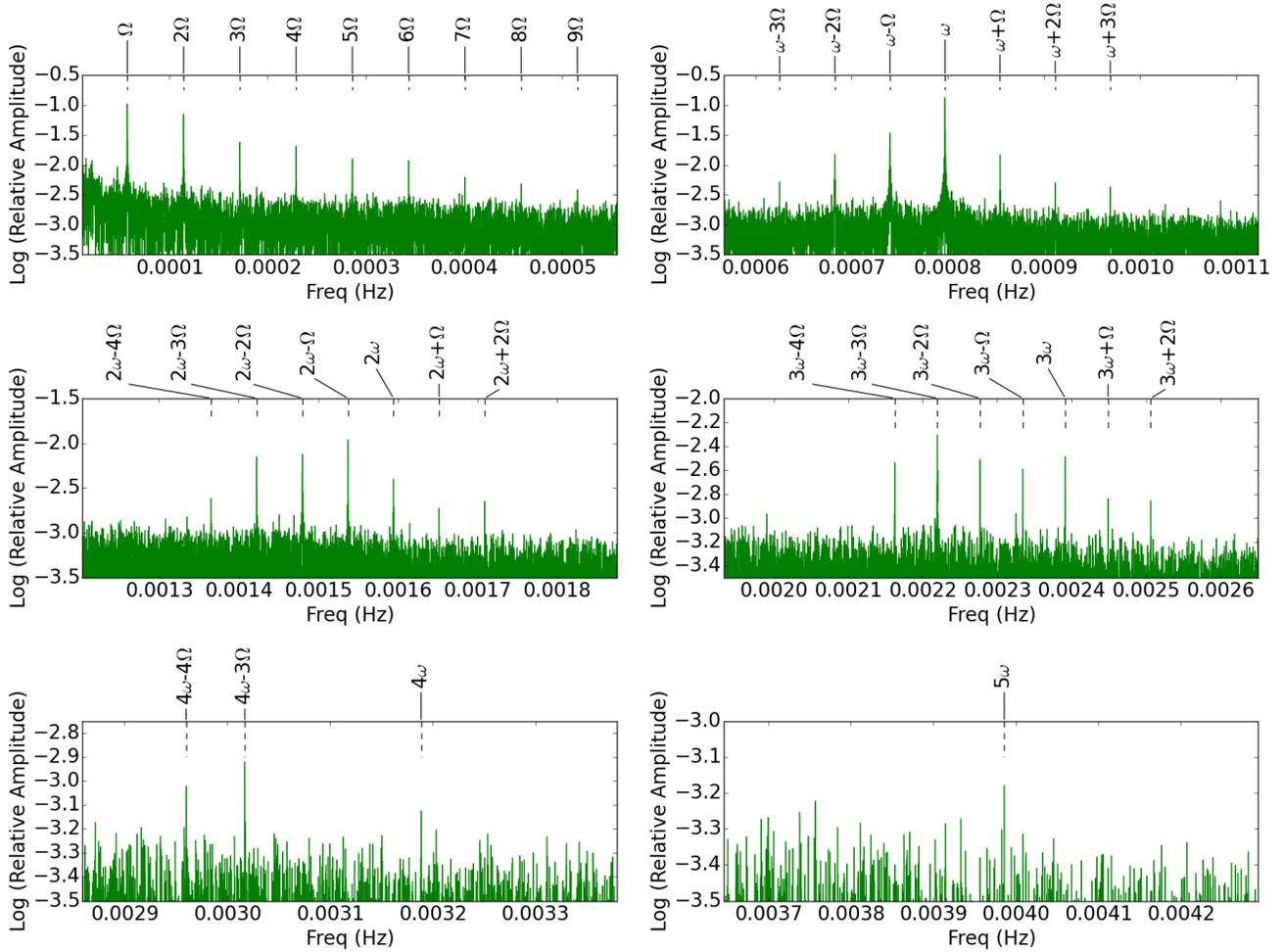}
	\caption{Several cuts of the amplitude spectrum, around the main/strongest signals in the light curve. The strongest signals are the orbital frequency $\Omega$ at 0.57265$\times10^{-4}$ Hz and the spin frequency $\omega$ at 7.972319(2)$\times10^{-4}$ Hz. A large number of sideband frequencies are also clearly detected and are identified as labelled.}
	\label{1minps}
\end{figure*}

The 1 min cadence data were subjected to a Lomb-Scargle Periodogram (\citealt{Lomb76}; \citealt{scargle82}) using the astroML python library \citep{astroML} to search for the spin period of the WD and the orbital period. A small section of the amplitude spectrum, around the main/strongest signal in the light curve, can be seen in Figure \ref{pssingle} and Figure \ref{1minps}. The top left panel of Figure \ref{1minps} shows the orbital frequency of 0.57265$\times10^{-4}$ Hz and 9 higher harmonics. The top right panel of Figure \ref{1minps} has a strong frequency peak at 7.972319(2)$\times10^{-4}$ Hz (P = 1254.3401(4) s). Also visible is the beat frequency of 7.40015(1)$\times10^{-4}$ Hz ($P_{b}$= 1351.329(2) s). Table \ref{frequency_tab} shows the frequency of all detected peaks, along with their amplitudes.

\begin{table}
	\centering
	\caption{Frequency peaks seen in Figure \ref{1minps}.}
	\begin{tabular}{r c c c}
		\hline
								& Frequency			& Period					& Relative Amplitude\\
								& Hz ($\times10^{-4}$)				& h				& \\
		\hline\hline
		$\Omega$					& 0.57265(4)			& 4.8508(4)			& 0.101(4)\\
		2$\Omega$				& 1.1456(8)			& 2.425(3)			& 0.071(3)\\
		3$\Omega$				& 1.7185(8)			& 1.6164(8)			& 0.0240(3)\\
		4$\Omega$				& 2.2912(8)			& 1.2123(4)			& 0.0208(3)\\
		5$\Omega$				& 2.8639(9)			& 0.9699(3)			& 0.0126(3)\\
		6$\Omega$				& 3.4369(8)			& 0.8082(2)			& 0.0118(2)\\
		7$\Omega$				& 4.007(1)			& 0.6928(2)			& 0.0061(2)\\
		8$\Omega$				& 4.5825(8)			& 0.6062(1)			& 0.0048(1)\\
		9$\Omega$				& 5.1547(9)			& 0.5389(1)			& 0.0038(1)\\
		\hline
								& 					& s				& \\
		\hline
		$\omega$					& 7.972319(2)		& 1254.3401(4)		& 0.1367(4)\\
		$\omega$-$\Omega$		& 7.40015(1)			& 1351.329(2)		& 0.0282(4)\\
		$\omega$+$\Omega$		& 8.54509(1)			& 1170.262(2)		& 0.0205(4)\\
		$\omega$-2$\Omega$		& 6.82652(1)			& 1464.876(2)		& 0.0222(4)\\
		$\omega$+2$\Omega$		& 9.1179(8)			& 1096.74(6)			& 0.0050(3)\\
		$\omega$-3$\Omega$		& 6.2541(8)			& 1599.0(2)			& 0.0052(3)\\
		$\omega$+3$\Omega$		& 9.6907(8)			& 1031.94(6)			& 0.0042(2)\\
		\hline
		2$\omega$				& 15.9443(8)			& 627.18(3)			& 0.0040(2)\\
		2$\omega$-$\Omega$		& 15.3721(9)			& 650.53(4)			& 0.0107(3)\\
		2$\omega$+$\Omega$		& 16.5177(9)			& 605.41(4)			& 0.00185(3)\\
		2$\omega$-2$\Omega$		& 14.7993(9)			& 675.71(4)			& 0.0075(3)\\
		2$\omega$+2$\Omega$		& 17.090(1)			& 585.12(4)			& 0.0023(2)\\
		2$\omega$-3$\Omega$		& 14.2267(8)			& 702.91(4)			& 0.0071(3)\\
		2$\omega$-4$\Omega$		& 13.6539(8)			& 732.39(4)			& 0.0024(2)\\
		\hline
		3$\omega$				& 23.9171(9)			& 418.11(2)			& 0.0032(2)\\
		3$\omega$-$\Omega$		& 23.3443(9)			& 428.37(2)			& 0.0025(2)\\
		3$\omega$+$\Omega$		& 24.490(1)			& 408.34(2)			& 0.0014(1)\\
		3$\omega$-2$\Omega$		& 22.7714(8)			& 439.15(2)			& 0.0030(2)\\
		3$\omega$+2$\Omega$		& 25.0628(8)			& 399.00(1)			& 0.0013(1)\\
		3$\omega$-3$\Omega$		& 22.1991(8)			& 450.47(2)			& 0.0048(4)\\
		3$\omega$-4$\Omega$		& 21.6258(8)			& 462.71(2)			& 0.0029(2)\\
		\hline
		4$\omega$				& 31.8880(9)			& 313.58(1)			& 0.00075(1)\\
		4$\omega$-3$\Omega$		& 30.1714(9)			& 331.44(1)			& 0.00120(3)\\
		4$\omega$-4$\Omega$		& 29.5987(8)			& 337.85(1)			& 0.00095(3)\\
		\hline
		5$\omega$				& 39.862(1)			& 250.87	(1)			& 0.00065(1)\\
		\hline
	\end{tabular}
	\label{frequency_tab}
\end{table}

The amplitude spectrum of FO Aqr in Figure \ref{pssingle} shows a break at $\sim10^{-3}$ Hz, where the slope of the continuum in the amplitude spectrum changes. This matches the break seen in other CVs such as MV Lyrae \citep{Scaringi2012}, V1504 Cyg, KIC 8751494 \citep{VandeSande2015}, BZ Cam, CM Del, KR Aur, RW Tri, UU Aqr, and V345 Pav \citep{Scaringi2015}, and is associated with the rms-flux relation. The high frequency flickering associated with the rms-flux relation is thought to originate from the inner parts of the accretion disk close to the surface of the WD \citep{Scaringi2012}. If this is true, then it is surprising that the frequency break in FO Aqr matches the frequency break in systems such as MV Lyrae, since the inner radius of the accretion disk in FO Aqr, which is truncated by the magnetic field, is much larger than in these non-magnetic systems.

\subsection{Phased light curve}
\cite{Hellier1989} proposed that the companion in FO Aqr eclipses the near side of the accretion disk and suggested an inclination between 65\degree and 75\degree , based on the recurring minimum present in the light curve when phased to the orbital period. We now consider if this minimum is present for the 69 days of observations.

\begin{figure}
	\includegraphics[width=80mm]{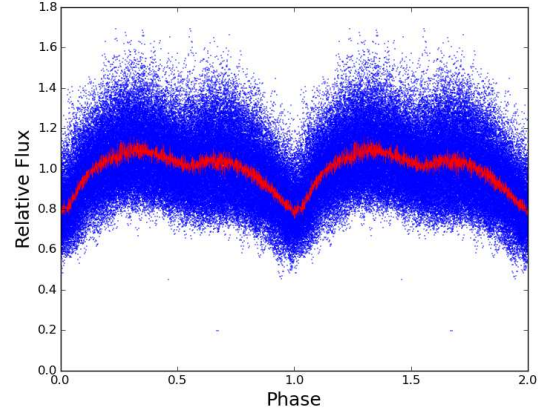}
	\caption{The phased light curve of FO Aqr using the orbital period. The 0 phase corresponds to the eclipse of the outer accretion disk as proposed by \cite{Hellier1989}. The red light curve uses a bin width of 60 points. The minimum rms of the light curve occurred at $\phi$=0}
	\label{foaqr_eclipse}
\end{figure}

We initially phased the light curve using the orbital period from Table \ref{frequency_tab}. Then, the phased light curve was binned, and the minimum of the light curve was set to $\phi$=0 (T$_{0}$(BJD)=2,456,982.2278(6)).
Figure \ref{foaqr_eclipse} shows this phased light curve. The eclipse of the outer accretion disk as proposed by \cite{Hellier1989} is clearly visible at $\phi$=0. The minimum rms of each bin in Figure \ref{foaqr_eclipse} occured at $\phi$=0, which is when the outer part of the accretion disk is being eclipsed.

\section{Variability of the Spin Signal}
Long-term monitoring of IPs usually reveals either a spin up or spin down of the WD, as the magnetic connection between the inner edge of the disk and a magnetic primary star exerts accretion torques on the magnetic primary (\citealt{Ghosh1978}; \citealt{Ghosh1979}). As mentioned previously, FO Aqr has a history of switching from spin up to spin down \citep{Patterson1998}. We now investigate whether we can detect variations in the spin signal of FO Aqr from the \textit{K2} data, and establish how the period has changed since the last reported value of 1254.284(16) s \citep{Andronov2005}.

T$_{0}$ for calculating the spin period was set to the maximum of one of the spin cycles located in the middle of the data set, at T$_{0}$(BJD)=2,457,012.4554(1).

\subsection{Variability of Spin Period} \label{VarSpinPeriod}
\begin{figure}
	\includegraphics[width=80mm]{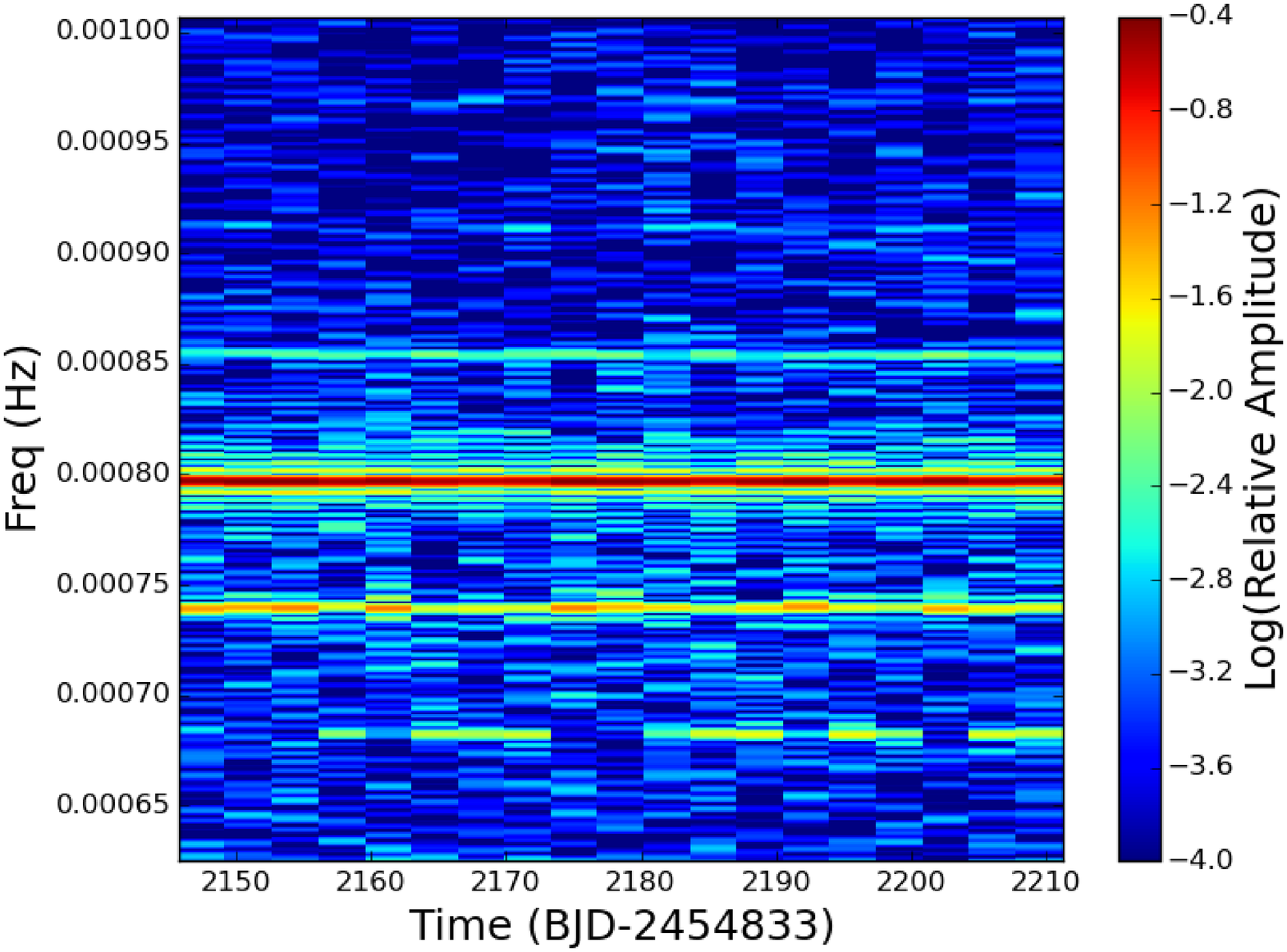}
	\includegraphics[width=80mm]{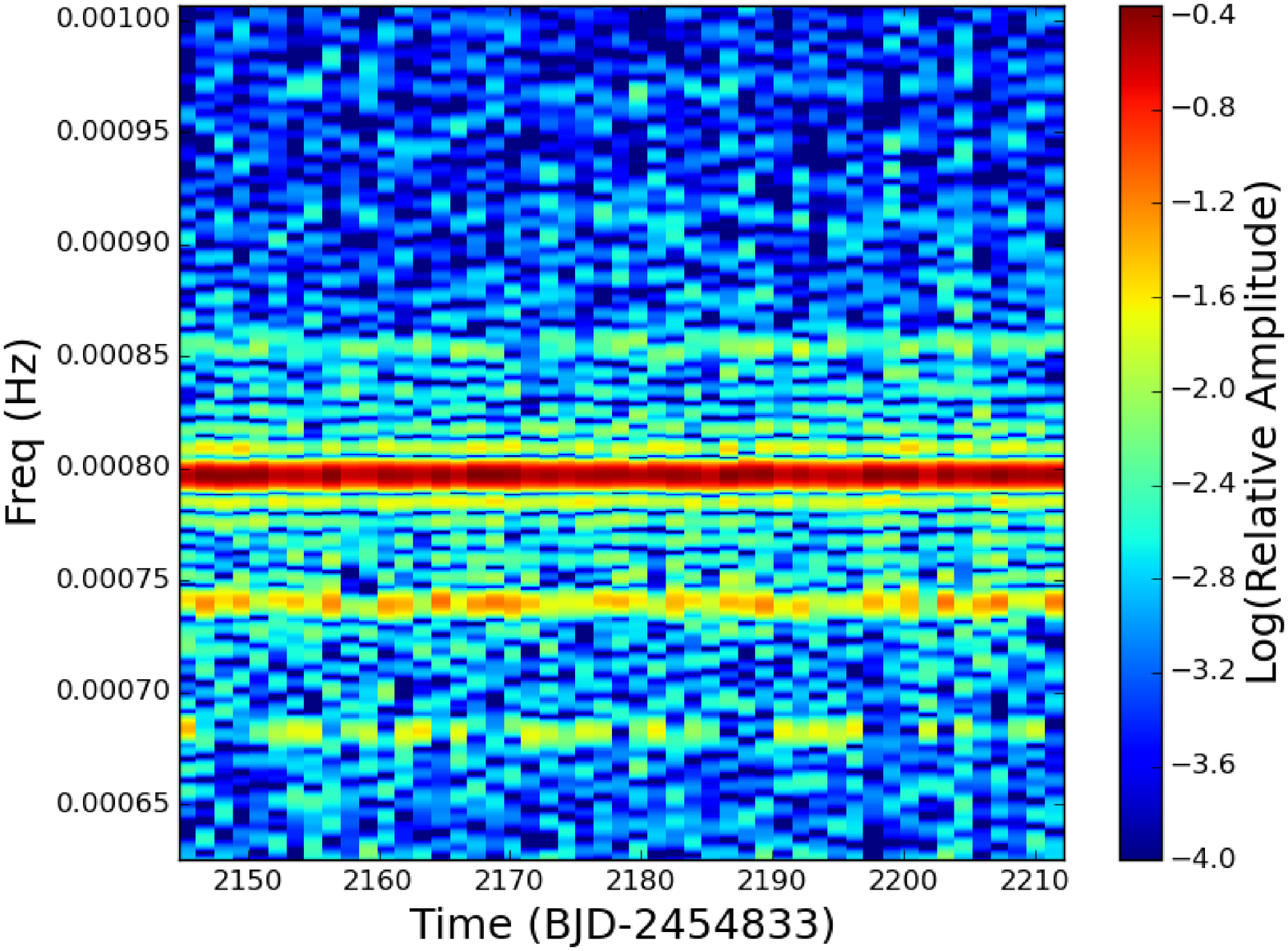}
	\caption{{\it Top Panel}: Running LS Periodogram for 5000 min bins. {\it Bottom Panel}: Same, but for 2000 min bins.}
	\label{runningps}
\end{figure}

In order to look for variations in the WD spin frequency, the light curve was split into segments and each segment was subjected to a Lomb Scargle Periodogram. The results of 2 different segment sizes (5000 points (3.5 days) per segment and 2000 points (1.4 days) per segment) are shown in Figure \ref{runningps}. The WD spin frequency is stable for both segment sizes, as is the beat frequency. The stability of the spin frequency would normally be associated with a stable mass transfer rate over the observations. However, as we will show in Section \ref{VarSpinPhase}, the observations were not sensitive to large variations in $\dot{m}$.

The power of the $\omega-2\Omega$ peak seems to vary with time, but the detectability of this peak is strongly dependent on the size of the data segment used to estimate the amplitude. Hence we attribute this to poor signal-to-noise, as opposed to a physical mechanism within the system. To better see this, consider the segments before T(BJD-2454833)=2190 in both the top and bottom plots. Using the 5000 point per bin data (top plot), the $\omega-2\Omega$ peak is clearly visible with strong power. However, using the 2000 point per bin data, this peak is lost in noise. This is also visible in reverse in the bins just after T(BJD-2454833)=2190, where $\omega-2\Omega$ is lost in noise in the 5000 point per bin data, while it is well detected in the 2000 point per bin data.

\begin{figure}
	\includegraphics[width=80mm]{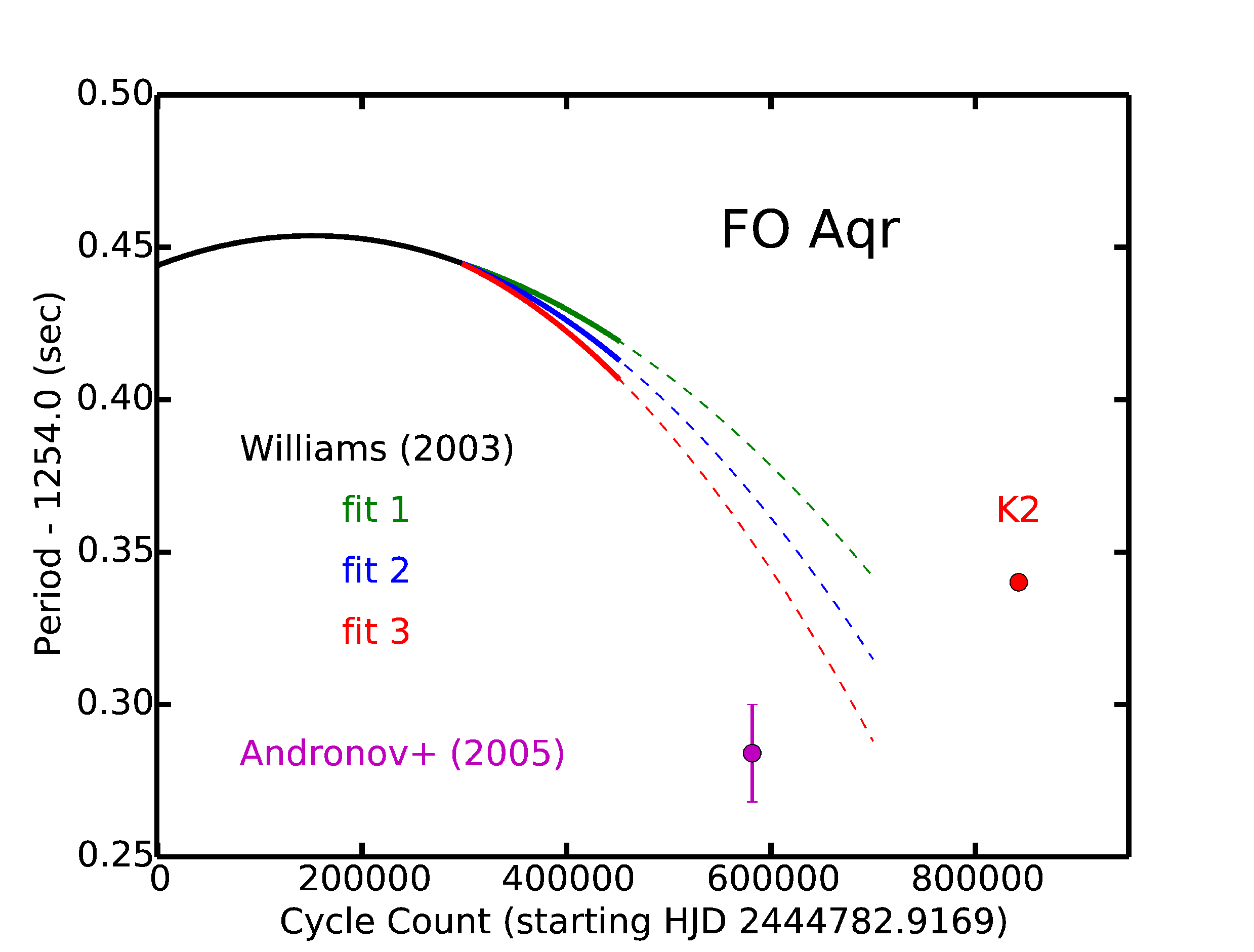}
	\caption{The observed spin period changes in FO Aqr (solid line), adopted from \cite{Williams2003} based on 20 years of data. The dashed lines show three possible 4th order extrapolations from \cite{Williams2003} caused by an uncertainty in the cycle count. The period measure by K2 is shown at the right and the errorbars in both period and cycle count are smaller than the size of the point. \cite{Andronov2005}  measure a spin period that is shorter than that seen by K2 suggesting that some time between 2005 and 2014, FO Aqr ceased spinning up, and began to spin down. 
}
	\label{williamspredict}
\end{figure}

The spin period detected by \textit{K2} was 1254.3401(4) s. This period is shorter than the previously recorded periods of 1254.4446(2) s (\citealt{Patterson1998}; \citealt{Kruszewski1998}) and 1254.4441(1) s \citep{Williams2003}, but longer than the period from 2004 of 1254.284(16) s \citep{Andronov2005}. Figure \ref{williamspredict} shows three 4th order fits to the data in \cite{Williams2003}, along with the period from \cite{Andronov2005} and our \textit{K2} period. This suggests that between 2005 and the beginning of the \textit{K2} observations, FO Aqr ceased spinning up, and began to spin down. Taking the period from the \textit{K2} observations and the period from \cite{Andronov2005} gives an estimate of the current $\dot{P}=2.0(5)\times10^{-10}$ s s$^{-1}$. \cite{Williams2003} estimated $\dot{P}$ to be $-4.1\times10^{-10}$ s s$^{-1}$ by 2015 based on their 3rd order ephemeris. However, the large cycle-count ambiguity between the \textit{K2} observations and the observations in \cite{Patterson1998}, \cite{Kruszewski1998} and \cite{Williams2003} prevents a direct linking of these datasets.

\subsection{Variability of Spin Phase} \label{VarSpinPhase}
There is no evidence of a significant $\dot{P}$ from our data alone in Section \ref{VarSpinPeriod}, so we now look for changes in the phase of the spin signal, which is a more sensitive technique than Section \ref{VarSpinPeriod}, to confirm the lack of a $\dot{P}$ in the Kepler data.

\begin{figure}
	\includegraphics[width=90mm]{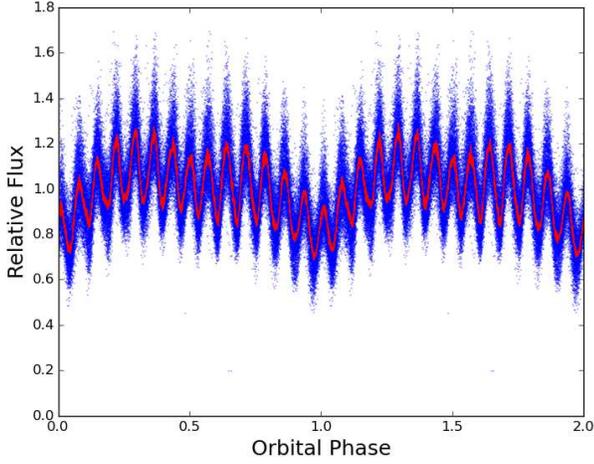}
	\includegraphics[width=80mm]{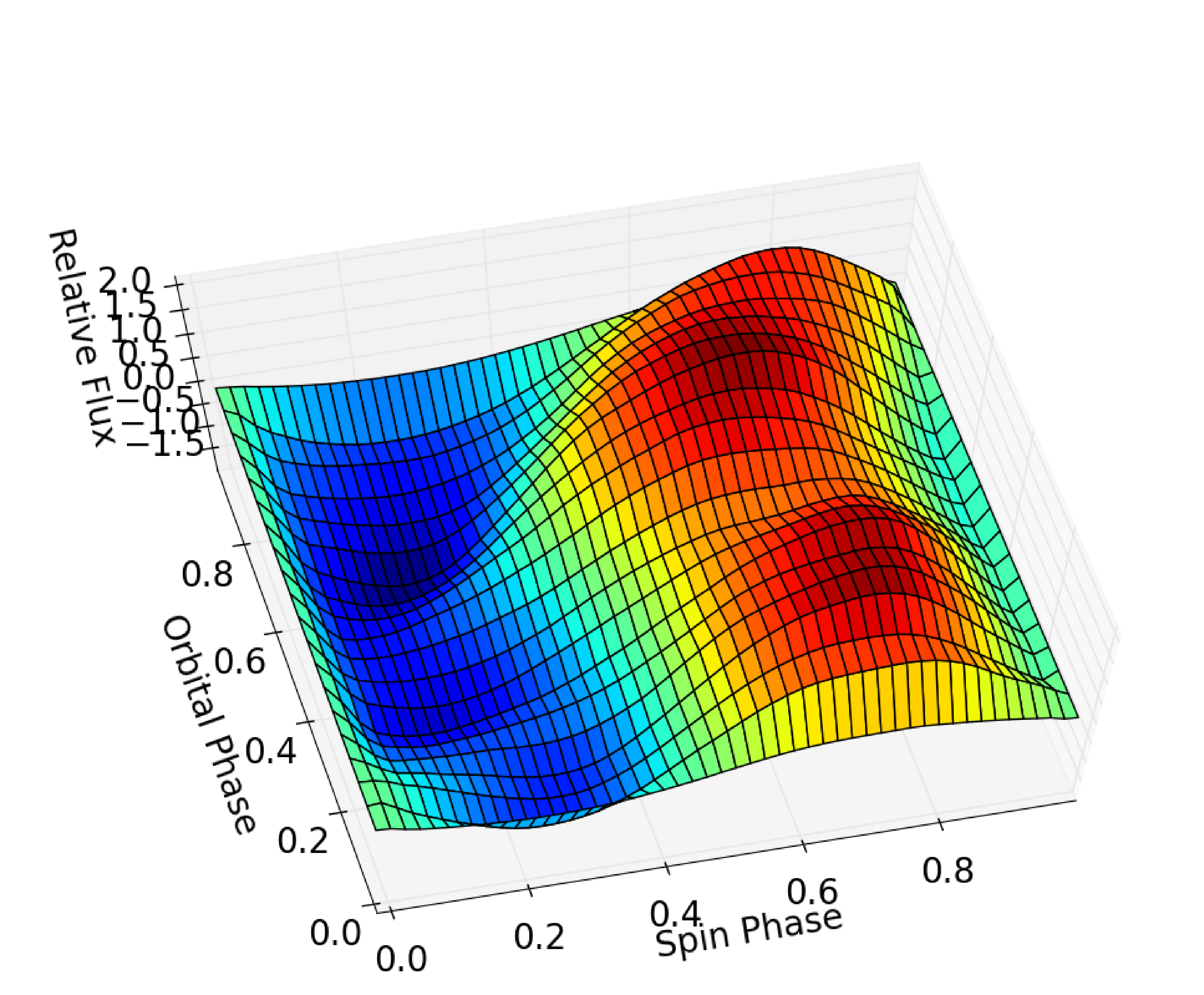}
	\caption{{\it Top Panel}: The phased light curve of FO Aqr, where the spin phase for each successive orbit has been shifted so that 0 spin phase of each orbit corresponds to 0 orbital phase. {\it Bottom Panel}: A 3-D representation of the spin signal as a function of the orbital phase. The peak of the spin signal arrives earliest near orbital phase 0.7 and latest just before eclipse.}
	\label{foaqr_phased}
\end{figure}

In order to investigate whether the spin signal of the WD varies over the orbital period, the light curve was phased using the orbital and spin periods. The orbital and spin phase was calculated for each data point in the light curve and then, to align the spin phases, the zero phase of the spin period closest to the zero phase of the orbital period was shifted so that the two zero phases overlapped. As there are several cycles of the spin signal within an orbital period of the binary, we can measure the zero phase of spin signal accurately and align it with the zero phase of the orbital period. This shows the variation due to the spin signal of the white dwarf within a single binary period. We now investigate whether there is a change in the spin phase over the length of the light curve, which will indicate a spin up or spin down of the white dwarf. The resulting phased light curve can be seen in the top panel of Figure \ref{foaqr_phased}, which shows how the amplitude and phase of the spin signal changes over the course of the orbital period. This is better seen in the 3-D plot in the bottom panel of Figure \ref{foaqr_phased}, which show that the amplitude of the spin signal is strongest at the orbital phase of 0.75 and 0.25, and also that the arrival time of the spin signal depends on the orbital phase. At orbital phase 0, the spin signal arrives when expected, but at orbital phase 0.5, the spin signal arrives earlier than expected. This means the phase of the spin signal changes over the orbital period.

\begin{figure}
	\includegraphics[width=80mm]{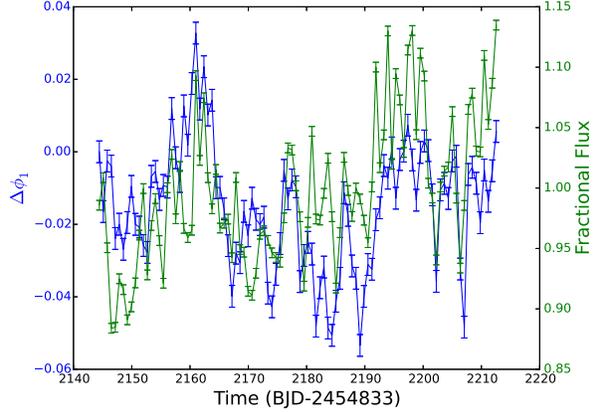}
	\caption{The change in phase of the spin signal (blue) and the change in flux relative to the 69 day average (green) versus time. The change in the phase of the spin signal is thought to be related to either the torque exerted on the WD or a change in the mass accretion rate.}
	\label{multisine_phase_plot}
\end{figure}

To investigate the change of the spin signal over the entire observation window, the data were split up into sections consisting of 1000 points each ($\sim 50$ spin periods per section). A multi-sine function of the form
\begin{equation}
\label{multisine}
F(t) = \sum_{n=1}^{4} A_{n} \sin \left( 2\pi\left(f_{n}t+\phi_{n}\right)\right)
\end{equation}
was fit to the white dwarf spin signal, after the orbital variation in the light curve had been removed. The values for $f_{1}$, $f_{2}$, $f_{3}$ and $f_{4}$, which are the spin frequency and the 3 strongest side band frequencies, were set at 7.972319$\times10^{-4}$ Hz, 7.40015$\times10^{-4}$ Hz, 8.54509$\times10^{-4}$ Hz and 6.82652$\times10^{-4}$ Hz respectively using the values from Table \ref{frequency_tab}, and each $A_{n}$ and $\phi_{n}$ was allowed to vary. Figure \ref{multisine_phase_plot} shows $\Delta\phi_{1}$, the change in phase of the spin signal, versus time along with the change in flux from FO Aqr.

Figure \ref{multisine_phase_plot} shows that $\Delta\phi_1$, the change in phase of the spin signal, correlates well with the observed change in flux from FO Aqr.
If this change in phase of the spin signal is due to the torque between the accretion disk and the WD, then the radius at which the accretion disk is truncated by the magnetic field can be estimated. 

First, assume the instantaneous accretion rate $\dot{m}$ results in a spin up of the WD 
\begin{equation}
\label{spinup}
I\frac{d\omega}{dt}=\dot{m}R^{2}_{K}\Omega_{K}-\alpha
\end{equation}
where $I$ is the moment of inertia of the WD, $R_{K}$ is the radius 
at the coupling point where the material connects to the magnetic field and $\Omega_{K}=\sqrt{GM/R^{3}_{K}}$ is the corresponding Keplerian angular velocity. $\alpha$ represents a constant drag on the magnetic field by the disc material. 
Assuming that the mean accretion rate, $\left\langle\dot{m}\right\rangle$, results in no net spin up or down of the WD, $\alpha$ can be expressed as $\alpha = \left\langle\dot{m}\right\rangle \sqrt{GMR_{K}}$, so that
\begin{equation}
\label{spinup1}
I\frac{d\omega}{dt}=\sqrt{GMR_{K}}\left(\dot{m}-\left\langle\dot{m}\right\rangle\right).
\end{equation}
The change in accretion rate above its mean value also results in an increase of the observed flux $f$. Let $\beta$ be a constant scaling factor, then $\dot{m}-\left\langle\dot{m}\right\rangle=\beta\Delta f$.
Finally, writing the spin frequency in terms of the phase change $\omega=\dot{\phi}$ and integrating twice, we find,
\begin{equation}
\label{spinup2}
\phi(t)=\phi(0)+\dot{\phi}(0)t+\beta\frac{\sqrt{GMR_{K}}}{I}\int_{0}^{t}\int_{0}^{t'} \Delta f(t'')  dt'' dt'
\end{equation}

$\Delta f$ here is the fractional flux shown in Figure~\ref{multisine_phase_plot}. We numerically integrated equation~\ref{spinup2} to determine $R_{K}$, i.e. the disc truncation radius required if the observed phase change was due to a spin-up torque. Using a canonical value for the WD mass in a CV ($M \sim 0.87 M_{\odot}$ (\citealt{Littlefair2008}; \citealt{Zorotovic2011})) and the average accretion rate for FO Aqr of $\left\langle\dot{m}\right\rangle \sim 1\times10^{-9} M_{\odot}$~yr$^{-1}$ calculated by \cite{Williams2003}, then we find that $R_{K}\approx10^{7}$ R$_{\odot}$. The binary separation in a $P_{\mbox{\small orb}} = 4.85$~h binary such as FO Aqr is $\approx$1.5 R$_{\odot}$, so the required $R_{K}$ is orders of magnitude larger than the binary separation of the system. As such, we attribute the change in the phase of the spin signal to a possible change in either the mass accretion rate or the accretion geometry, as opposed to a genuine variation in the spin phase of the WD.

\subsection{Modulation of the Spin and Beat Frequencies}
\begin{figure}
	\includegraphics[width=80mm]{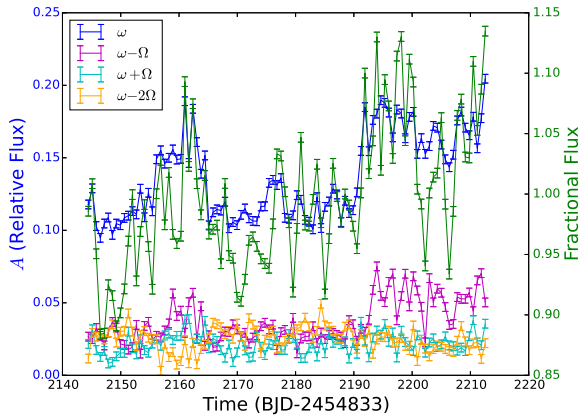}
	\caption{The change in amplitudes, measured in relative flux, for the spin signal (blue) and its strongest sidebands (magenta, cyan and orange) for the multi sine curve fit to the spin signal and the change in flux relative to the 69 day average (green) versus time. The change in the amplitude of the $\omega-\Omega$ peak (magenta) is thought to be related to a change in the accretion rate in the system, as it is correlated with the increased flux and increased amplitude of the spin signal.}
	\label{multisine_plot}
\end{figure}

Figure \ref{multisine_plot} shows the amplitudes of the various components for each section of data, found in the same way that $\Delta\phi$ was found in the previous section, along with the change in flux relative to the 69 day average for each section of data from the uncorrected \textit{Kepler} light curve.

The amplitude variations seen in the spin and beat signals strongly correlate with the change in flux from the system, as seen in Figure \ref{multisine_plot}. This suggests the mass accretion rate in this system is not stable over a timescale of $\sim50$ spin periods. However, this is due to the increased sensitivity of this method of fitting sine waves with a fixed period (vs a running periodogram) to small changes in the mass accretion rate.

For the first 50 days of observations, the amplitudes of the $\omega-\Omega$ and the $\omega+\Omega$ signals were similar, and the flux in the system increased and decreased in line with the changing amplitude of the spin signal. The easiest way to have equal power in both of these sideband frequencies was if the accretion geometry was a simple disk, without accretion disk overflow, as proposed by \cite{Norton1992}. After day 2192 (BJD-245833), the amplitude of the $\omega-\Omega$ peak began to increase, while the amplitude of $\omega+\Omega$ remained stable. This suggests an increase in the accretion rate in the system at this time, which caused more optical light to be reflected by the source of the beat signal in the system (either the hot spot or the heated side of the companion).The increased accretion rate is supported by the sustained increase in the amplitude in the spin frequency and the overall flux.

\section{Conclusion}
The 69 day light curve of FO Aqr taken by the \textit{Kepler} \textit{K2} mission displays the power of observing an IP continuously for long periods of time. The \textit{K2} data gave an accurate spin period of the WD and determined the varying power in the sidebands of the amplitude spectrum over 69 days of continuous observation. It provided comparable accuracy to ground based observations which took place over decades. This is a true testament to the power of \textit{K2} in studying the time varying mechanisms that are present in magnetically accreting objects.

The eclipse of the outer accretion disk proposed by \cite{Hellier1989} is consistently seen throughout the 69 day light curve. The minimum of the rms is seen at orbital phase zero also, in line with the outer part of the accretion disk being eclipsed. There is no evidence for a precessing disk in the system, which would have shown up as a period longer than the orbital period in the amplitude spectrum.  

The spin period has increased to 1254.3401(4) s since the last reported value from 2005, when FO Aqr was shown to be spinning up. It appears this spin up has ceased, and further short observations over the coming year should be carried out to confirm that FO Aqr is now in a state of spinning down, and to allow for an accurate value of $\dot{P}$ to be determined using our T$_{0}$ and spin period.

The \textit{K2} observations will help to establish an unambiguous cycle count for future observations of FO Aqr, which will be important in determining if the current spin down in FO Aqr continues. Future \textit{K2} observations of other IPs may help us to probe changes in their spin frequencies on time scales as short as $\sim$2.5 months.

\section*{Acknowledgments}
This work made use of PyKE \citep{2012ascl.soft08004S}, a software package for the reduction and analysis of \textit{Kepler} data. This open source software project is developed and distributed by the NASA \textit{Kepler} Guest Observer Office. PS acknowledges support from NSF grant AST-1514737. MRK and PG acknowledge support from the Naughton Foundation and the UCC Strategic Research Fund. EB, TRM and DS are supported by the STFC under grant ST/L000733. BTG is supported by the ERC Advanced Grant n. 320964 (WDTracer). ZD is supported by CAS ``Light of West China” Program.'' We would like to thank the anonymous referee for their comments in improving this paper.

\bibliographystyle{mn2e}
\bibliography{foaqr_kennedy_apr_16.bib}

\label{lastpage}
\end{document}